\documentclass[12pt]{article}
\usepackage{graphicx}
\begin{document}
\begin{center}
{\large\bf Modified Gravitational Theory and the Gravity Probe-B Gyroscope
Experiment} \vskip 0.3 true in {\large J. W. Moffat}
\vskip 0.3 true in
{\it The Perimeter Institute for Theoretical Physics, Waterloo, Ontario,
N2J 2W9, Canada}
\vskip 0.3 true in and
\vskip 0.3 true in {\it Department
of Physics, University of Toronto, Toronto, Ontario M5S 1A7, Canada}
\end{center}
\begin{abstract}%
A possible deviation from the precession of the Gravity Probe-B gyroscope
predicted by general relativity is obtained in the
nonsymmetric gravity theory. The time delay of radio signals emitted
by spacecraft at planetary distances from the Sun, in nonsymmetric gravity
theory is the same as in general relativity. A correction to the
precession of the gyroscope would provide an experimental
signature for the Gravity Probe-B gyroscope experiment. The Lense-Thirring
frame-dragging effect is predicted to be the same as in GR. \end{abstract}
\vskip 0.2 true in e-mail: jmoffat@perimeterinstitute.ca


\section{Introduction}

The Gravity Probe-B spacecraft was launched on April 19, 2004, at
Vandenberg Air Force Base, California, and is expected to produce
experimental results that test the prediction of the relativistic
precession of the onboard gyroscope. The polar orbit will be at a height of
$640\,{\rm km}$ passing over every pole at $48.75\,{\rm min}$.

A gravitational theory explanation of the acceleration of the
expansion of the universe~\cite{Perlmutter,Riess,Spergel} and the
observed flat rotation curves of galaxies was
proposed~\cite{Moffat,Moffat2}, based on the nonsymmetric gravitational
theory (NGT)~\cite{Moffat3,Moffat4,Moffat5}. The equations of
motion of a test particle in a static, spherically symmetric
gravitational field modifies Newton's law of acceleration for
weak fields. It was shown~\cite{Moffat6} that the modified acceleration can
account for the anomalous acceleration observed for the Pioneer
10 and 11 spacecraft~\cite{Anderson}.

In the following, we shall calculate a possible deviation from the
predicted precession of the gyroscope obtained from general relativity (GR)
using the test particle equations of motion in NGT. A previous calculation
of such a deviation was based on an earlier version of
NGT~\cite{Moffat3,Brownstein}. In 1995, a new version of NGT was published
in which the weak field linear approximation of the field equations was
shown to be stable and the Hamiltonian was bounded from below leading to a
stable vacuum state~\cite{Moffat4,Moffat5}. We shall calculate the
gyroscope motion from the 1995 version of NGT, using the static,
spherically symmetric solution of the vacuum field equations and the
equations of motion of test particles~\cite{Moffat2}.

The gravitational constant at infinity is defined to be
\begin{equation}
\label{renormG}
G\equiv G_{\infty}=G_0\biggl(1+\sqrt{\frac{M_0}{M}}\biggr),
\end{equation}
where $G_0$ is Newton's gravitational constant, $M$ is the mass of a
physical source and $M_0$ is a positive parameter. In the NGT
phenomenology, there is also a range parameter $r_0=1/\mu$, which is
distance dependent in that it increases as we go from small to large
gravitationally bound systems.

One important result derived from NGT is that photons move along null
geodesics that predict the same bending of light and time delay effects
for the Sun as in GR. The time delay of radio signals passing the limb of
the Sun is shown in NGT to be the same for weak gravitational fields as in
GR. This means that spacecraft measurements of the time delay of
radio signals that determine the post-Newtonian parameter $\gamma$ to be
equal to unity to a high precision, do not prove that it is a foregone
conclusion that the Gravity Probe-B experiment for the measurement of the
gyroscope precession should agree with GR. The correction
to the GR predicted value of the gyroscope's precession, obtained
from the NGT equations of motion of a spinning particle, could be detected
by the sensitive Gravity Probe-B experiment. The frame-dragging effect
predicted by NGT is the same as in GR.

\section{The Equations of Motion of a Spinning Particle}

The equations of motion of a test particle can be obtained from the NGT
conservation laws~\cite{Legare,Moffat2}:
\begin{equation}
\frac{1}{2}(g_{\mu\rho}{{\bf T}^{\mu\nu}}_{,\nu}+g_{\rho\mu}{{\bf
T}^{\nu\mu}}_{,\nu}) +[\mu\nu,\rho]{\bf T}^{\mu\nu}=0, \end{equation}
 where
\begin{equation}
[\mu\nu,\rho]=\frac{1}{2}(g_{\mu\rho,\nu}+g_{\rho\nu,\mu}-g_{\mu\nu,\rho})
\end{equation}
and ${\bf T}^{\mu\nu}=\sqrt{-g}T^{\mu\nu}$. The test particle source ${\bf
T}^{\mu\nu}$ vanishes outside a narrow tube in four-dimensional spacetime
that surrounds the coordinate point $X^\mu$ of the particle.
The motion of a monopole particle is obtained assuming that the dipole and
higher moments of ${\bf T}^{\mu\nu}$ vanish.

We shall consider the motion of the test particle coupled to a skew
symmetric source. This yields the equations of motion~\cite{Moffat2}:
\begin{equation} \label{geodesic}
\frac{du^\mu}{ds}+\left\{{\mu\atop\alpha\beta}\right\}u^\alpha
u^\beta=s^{(\mu\sigma)}f_{[\sigma\nu]}u^\nu,
\end{equation}
where $u^\mu=dx^\mu/ds$ is the 4-velocity of a particle and
$s$ is the proper time along the path of the particle.
Moreover, $s^{(\mu\alpha)}g_{(\nu\alpha)}={\delta^\mu}_\nu$ and
\begin{equation}
\left\{{\mu\atop\alpha\beta}\right\}=
\frac{1}{2}s^{(\mu\sigma)}(\partial_\beta g_{(\alpha\sigma)}
+\partial_\alpha g_{(\beta\sigma)}-\partial_\sigma g_{(\alpha\beta)}).
\end{equation}
The notation is the same as
in refs.~\cite{Moffat,Moffat2,Moffat4,Moffat5}.

The spin angular momentum is defined by
\begin{equation}
\label{spin}
S^{\mu\nu}=\int d^3x[(x^\mu-X^\mu){\bf T}^{\nu0}-(x^\nu-X^\nu){\bf
T}^{\mu0}],
\end{equation}
where $S^{\mu\nu}$ has the transformation properties of a
tensor. We assume that the test particle is a pole-dipole particle,
whereby the pole and dipole contributions of the integrated test
particle source ${\bf T}^{\mu\nu}$ do not vanish. The equations of motion
of the spin are found to be
\begin{equation}
\label{equationmotion}
\frac{DS^{\mu\nu}}{Ds}+u^\mu u_\gamma\frac{DS^{\nu\gamma}}{Ds}-u^\nu
u_\gamma\frac{DS^{\mu\gamma}}{Ds}=0,
\end{equation}
where
\begin{equation}
\label{covariantspin}
\frac{DS^{\mu\nu}}{Ds}\equiv
\frac{dS^{\mu\nu}}{ds}+\left\{{\mu\atop\alpha\beta}\right\}
S^{\alpha\nu}u^\beta+\left\{{\nu\atop\alpha\beta}\right\}S^{\mu\alpha}u^\beta.
\end{equation}

We shall use the condition
\begin{equation}
\label{spincondition}
S^{\mu\nu}u_\nu=0,
\end{equation}
which leads to $u^i=0\,(i=1,2,3)$ in the rest frame of the test particle
and the condition $S^{i0}=0$.
From (\ref{spincondition}), we get
\begin{equation}
\label{spinmotion}
\frac{DS^{\mu\nu}}{Ds}u_\nu+S^{\mu\nu}\frac{Du_\nu}{Ds}=0.
\end{equation}
Substituting (\ref{spinmotion}) into (\ref{equationmotion}), we get the
equations of motion for the spin
\begin{equation}
\label{spinequation}
\frac{DS^{\mu\nu}}{Ds}=(u^\mu S^{\nu\alpha}-u^\nu
S^{\mu\alpha})\frac{Du_\alpha}{Ds},
\end{equation}
where we used $s^{(\mu\nu)}u_\mu u_\nu=1$. We have from
Eq.(\ref{geodesic}) that
\begin{equation}
\label{uequation}
\frac{Du_\mu}{Ds}=f_{[\mu\nu]}u^\nu,
\end{equation}
where
\begin{equation}
\frac{Du_\mu}{Ds}=\frac{du_\mu}{ds}-
\left\{{\rho\atop\mu\beta}\right\}u^\beta u_\rho.
\end{equation}
Substituting
(\ref{uequation}) into (\ref{spinequation}), we get
\begin{equation}
\label{spinequation2}
\frac{DS^{\mu\nu}}{Ds}=(u^\mu S^{\nu\alpha}-u^\nu
S^{\mu\alpha})f_\alpha,
\end{equation}
where $f_\alpha=f_{[\alpha\sigma]}u^\sigma$ and
\begin{equation}
\label{orthogonal}
u^\alpha f_\alpha=0.
\end{equation}

In NGT the $f_\alpha$ acts as an external force that makes $DS^{\mu\nu}/Ds$
non-vanishing and this will affect the motion of the gyroscope orbiting
Earth.

\section{Experimental Prediction for the Gyroscope}

From (\ref{spinequation2}) we obtain
\begin{equation}
\label{spatial}
\frac{DS^{ij}}{Ds}=(u^iS^{j\alpha}-u^jS^{i\alpha})f_\alpha.
\end{equation}
We define the spatial vector ${\bf S}$ in rectangular coordinates
\begin{equation}
{\bf S}=(S^{23},S^{31},S^{12})=(S^1,S^2,S^3).
\end{equation}
Here, ${\bf S}$ is the spin angular momentum vector with respect to the
point ${\bf r}=(X^1,X^2,X^3)$.  Eq.(\ref{orthogonal}) tells us that the
terms in (\ref{spatial}) that involve $f_i$ are order
$v\vert{\bf S}\vert\vert {\bf f}\vert$, while the terms involving $f_0$ are
of order $v^2\vert{\bf S}\vert\vert{\bf f}\vert$ and can be neglected.
Moreover, we have that $f_i\sim -f^i$. For a slowly moving gyroscope,
we find
\begin{equation}
\label{gyroscopemotion}
\frac{D{\bf S}}{Dt}={\bf S}({\bf v}\cdot{\bf f})-{\bf f}({\bf v}
\cdot{\bf S}).
\end{equation}

For large values of $r$ the metric line element is given by
\begin{equation}
\label{lineelement}
ds^2\equiv
g_{(\mu\nu)}dx^\mu
dx^\nu=\biggl(1-\frac{2M}{r}\biggr)dt^2-\biggl(1+\frac{2M}{r}\biggr)
(dx^2+dy^2+dz^2), \end{equation}
where $r=(x^2+y^2+z^2)^{1/2}$. The non-vanishing Christoffel symbols are
\begin{equation}
\left\{{i\atop 00}\right\}=-\frac{1}{2}s^{(ik)}g_{00,k},\quad\quad
\left\{{0\atop i0}\right\}=\frac{1}{2}s^{00}g_{00,i},
$$ $$
\left\{{j\atop ik}\right\}
=\frac{1}{2}s^{(jl)}(g_{(il),k}+g_{(lk),i}-g_{(ik),l}).
\end{equation}

From (\ref{covariantspin}) and the left-hand side of (\ref{spatial}), we
have
\begin{equation}
\frac{D{\bf S}}{dt}=\frac{d{\bf S}}{dt}-\frac{2M}{r^3}({\bf r}\cdot{\bf
v}){\bf S}-\frac{M}{r^3}({\bf r}\times{\bf v})\times{\bf S}
-\frac{M}{r^3}({\bf r}\times{\bf W}),
\end{equation}
where we define
\begin{equation}
{\bf W}=(S^{10},S^{20},S^{30}).
\end{equation}
We have
\begin{equation}
({\bf r}\times{\bf v})\times{\bf S}={\bf v}({\bf r}\cdot{\bf S})-{\bf
r}({\bf v}\cdot{\bf S}).
\end{equation}
We obtain from (\ref{spincondition}) in isotropic coordinates
\begin{equation}
S^{10}\sim \biggl(1+\frac{4M}{r}\biggr)(v_yS_z-v_zS_y),
\end{equation}
which gives
\begin{equation}
{\bf W}\sim \biggl(1+\frac{4M}{r}\biggr)({\bf v}\times{\bf S})
\sim{\bf v}\times{\bf S}.
\end{equation}
We now obtain
\begin{equation}
{\bf r}\times{\bf W}\sim{\bf r}\times({\bf v}\times{\bf S})=
{\bf v}({\bf r}\cdot{\bf S})-{\bf S}({\bf r}\cdot{\bf v}).
\end{equation}
To our order of approximation, we have
\begin{equation}
\frac{D{\bf S}}{dt}=\frac{d{\bf S}}{dt}-\frac{M}{r^3}[{\bf S}({\bf
r}\cdot{\bf v})+2{\bf v}({\bf r}\cdot{\bf S})
-{\bf r}({\bf v}\cdot{\bf S})].
\end{equation}
From (\ref{gyroscopemotion}), we finally have
\begin{equation}
\frac{d{\bf S}}{dt}=\frac{M}{r^3}[{\bf S}({\bf r}\cdot{\bf v})
+2{\bf v}({\bf r}\cdot{\bf S})-{\bf r}({\bf v}\cdot{\bf S})]
+{\bf S}({\bf v}\cdot{\bf f})-{\bf f}({\bf v}\cdot{\bf S}).
\end{equation}

For weak gravitational fields, we have the equation of
motion~\cite{Moffat2}:
\begin{equation}
\label{slowmotion}
\frac{d{\bf v}}{dt}=-\biggl[\frac{M{\bf
r}}{r^3}-\sigma\frac{\exp(-r/r_0){\bf
r}}{r^3}\biggl(1+\frac{r}{r_0}\biggr)\biggr],
\end{equation}
where
\begin{equation}
\label{sigma}
\sigma=\frac{\lambda sM^2}{3r_0^2}.
\end{equation}
Here $r_0=1/\mu$ is the range parameter associated with the ``mass'' of the
skew field $g_{[\mu\nu]}$, and $\lambda$ and $s$ denote
the coupling strength of a test particle to matter and the coupling
strength of the skew field $g_{[\mu\nu]}$, respectively.

Let us consider the expansion for $r\ll r_0$:
\begin{equation}
\exp(-r/r_0)\biggl(1+\frac{r}{r_0}\biggr)
=1-\frac{1}{2}\biggl(\frac{r}{r_0}\biggr)^2+\frac{1}{3}\biggl(\frac{r}{r_0}\biggr)^3-...
\end{equation}
Then, (\ref{slowmotion}) gives for $r\ll r_0$:
\begin{equation}
\frac{d{\bf v}}{dt}\sim -\biggl[\frac{M{\bf
r}}{r^3}-\frac{\sigma{\bf r}}{r^3}
\biggl(1-\frac{1}{2}\biggl(\frac{r}{r_0}\biggr)^2\biggr)\biggr] \sim
-\biggl(\frac{M^*}{r^3}\biggr){\bf r},
\end{equation}
where
\begin{equation}
M^*=M-\sigma.
\end{equation}

We shall now transform to the rest frame of the gyroscope. This
yields~\cite{Schiff,Brownstein}:
\begin{equation}
{\bf S}_{\rm rest}=\biggl(1+\frac{2M}{r}\biggr){\bf S}
-\frac{1}{2}v^2{\bf S}+\frac{1}{2}{\bf v}({\bf v}\cdot{\bf S}).
\end{equation}
Differentiating this with respect to time gives the time dependence of the
spin vector in the rest frame of the gyroscope to lowest order:
\begin{equation}
\label{gyrospin}
\frac{d{\bf S}_{\rm rest}}{dt}
=\biggl(\frac{3M}{2r^3}-\frac{\sigma}{r^3}\biggr)[{\bf v}({\bf
r}\cdot{\bf S}_{\rm rest})-{\bf r}({\bf v}\cdot{\bf S}_{\rm rest})].
\end{equation}
Eq.(\ref{gyrospin}) can be written as
\begin{equation}
\frac{d{\bf S}_{\rm rest}}{dt}
=\biggl(\frac{3M}{2r^3}-\frac{\sigma}{r^3}
\biggr)({\bf r}\times{\bf v})\times{\bf S}_{\rm rest},
\end{equation}
or, in the form
\begin{equation}
\frac{d{\bf S}_{\rm rest}}{dt}={\bf\Omega}\times{\bf S}_{\rm rest},
\end{equation}
where
\begin{equation}
\label{omegaeq}
{\bf\Omega}=\biggl(\frac{3M}{2r^3}-\frac{\sigma}{r^3}\biggr)({\bf
r}\times{\bf v}).
\end{equation}

Let us take the gyroscope's orbit to be a circle of radius $r$ with unit
normal orbital angular momentum vector ${\hat{\bf J}}$. The gyroscope's
velocity is given by
\begin{equation}
{\bf v}=-\biggl(\frac{M^*}{r^3}\biggr)^{1/2}({\bf r}\times{\hat{\bf J}}).
\end{equation}
By using the relations
\begin{equation}
{\bf r}\times({\bf r}\times{\hat{\bf J}})=({\bf r}\cdot{\hat{\bf J}}){\bf
r}-({\bf r}\cdot{\bf r}){\hat{\bf J}}=-({\bf r}\cdot{\bf r}){\hat{\bf J}}
=-r^2{\hat{\bf J}},
\end{equation}
we obtain
\begin{equation}
{\bf\Omega}=\biggl(\frac{3M}{2r^3}
-\frac{\sigma}{r^3}\biggr)\biggl(\frac{M}{r^3}
-\frac{\sigma}{r^3}\biggr)^{1/2}r^2{\hat{\bf J}}
\sim \biggl(\frac{3M}{2}-\sigma\biggr)\frac{M^{1/2}}{r^{5/2}}\hat{\bf J}.
\end{equation}

For Earth, we choose
\begin{equation}
\label{earthvalue}
\sqrt{(M_0)_{\oplus}}\le 1.62\times 10^{-7}\,\sqrt{M_{\oplus}}
\end{equation}
and from this value and (\ref{renormG}) we see that $G\equiv
G_{\infty}\sim G_0$.

The precession rate in NGT to the lowest order of
approximation, averaged over a revolution for the Earth is given by (we
reinstate $G$ and $c$):
\begin{equation}
\label{NGTdeviation} \langle\vert{\bf\Omega}\vert\rangle
=B_{\oplus}\frac{(G_0M_{\oplus})^{1/2}}{c^2r^{5/2}}
=B_{\oplus}
\frac{(G_0M_{\oplus})^{1/2}}{c^2R_{\oplus}^{5/2}}
\biggl(\frac{R_{\oplus}^{5/2}}{r^{5/2}}\biggr),
\end{equation}
where
\begin{equation}
\label{Bequation}
B_{\oplus}=\frac{3}{2}G_0M_{\oplus}-\sigma_{\oplus}.
\end{equation}
Moreover, $M_{\oplus}$ and $R_{\oplus}$ denote the mass and the
equatorial radius of Earth, respectively, and we have
\begin{equation}
\label{sigmaconstant}
\sigma_{\oplus}=
\frac{\lambda_{\oplus} sG_0^2M^2_{\oplus}}{3c^2(r_0)^2_{\oplus}}.
\end{equation}

Setting $\sigma_{\oplus}=0$, we
get for the GR prediction
\begin{equation}
\langle\vert{\bf\Omega}\vert\rangle_{\rm GR}
=\frac{3(G_0M_{\oplus})^{3/2}}{2c^2R_{\oplus}^{5/2}}
\biggl(\frac{R_{\oplus}}{r}\biggr)^{5/2}
=8.40\biggl(\frac{R_{\oplus}}{r}\biggr)^{5/2}\,\,{\rm
arcsec}/{\rm yr}.
\end{equation}
For the Gravity Probe-B drag-free
spacecraft at a height of $h=640\,{\rm km}$ above Earth, we obtain for GR:
\begin{equation}
\langle\vert{\bf\Omega}\vert\rangle_{\rm GR}=6.61\,{\rm arcsec}/{\rm yr}.
\end{equation}

We see from (\ref{NGTdeviation}) and (\ref{Bequation}) that the correction
to GR reduces the gyroscope's precession by the amount of the value of
$\sigma_{\oplus}$. We expect to satisfy the bound
\begin{equation}
\sigma_{\oplus} < \frac{3G_0M_{\oplus}}{2},
\end{equation}
which yields from (\ref{sigmaconstant}):
\begin{equation}
\lambda_{\oplus}
< \frac{9c^2(r_0)_{\oplus}^2}{2sG_0M_{\oplus}}.
\end{equation}

\section{Frame Dragging Effect and Time Delay Observations}

In NGT, we can include the effect of the Earth's rotation to first order
in the metric in rectangular isotropic coordinates~\cite{Lense}:
\begin{equation}
g_{(\mu\nu)}=\left(\matrix{
-(1+2M/r)& 0 & 0 & -2{\bf J}y/r^3\cr
0 & -(1+2M/r) &  0 & 2{\bf J}x/r^3 \cr
0 & 0 & -(1+2M/r) & 0\cr
-2{\bf J}y/r^3 & 2{\bf J}x/r^3 & 0 & 1-2M/r\cr}
\right).
\end{equation}
The precession due to the Lense-Thirring effect, which arises from the
off-diagonal elements of the metric $g_{(\mu\nu)}$, is identical to GR. The
inclusion of the Earth's rotation has no effect to this order on the
diagonal elements of the metric.

The motion of massless photons is determined by the geodesic
equation~\cite{Moffat2}:
\begin{equation}
\label{nullgeodesic}
\frac{du^\mu}{ds}+\left\{{\mu\atop\alpha\beta}\right\}u^\alpha
u^\beta=0.
\end{equation}
To the order of approximation that we are considering,
the total light time of travel of a radio signal emitted from a spacecraft
in the solar, barycentric system is given by~\cite{Anderson}:
\begin{equation}
\label{delaytime}
t_2-t_1=\frac{r_{12}}{c}+\frac{2G_0M_{\odot}}{c^3}
\ln\biggl[\frac{r_{1\odot}+r_{2\odot}+r_{12\odot}}{r_{1\odot}+r_{2\odot}
-r_{12\odot}}\biggr]+\sum_i\frac{2G_0M_i}{c^3}\ln\biggl[\frac{r_{1i}+r_{2i}
+r_{12i}}{r_{1i}+r_{2i}-r_{12i}}\biggr],
\end{equation}
where $M_i$ is the mass of a planet, an outer
planetary system, or the Moon. Moreover, $r_{1\odot}$, $r_{2\odot}$ and
$r_{12\odot}$ are the heliocentric distances to the point of the radio
signal emission on Earth, to the point of signal reflection at the
spacecraft, and the relative distance between these two points,
respectively. The $r_{1i}$, $r_{2i}$ and $r_{12i}$ are distances relative
to an ith body in the solar system, and $t_1$ refers to the transmission
time at a tracking station on Earth, and $t_2$ refers to the reflection
time at the spacecraft. This is the same result  as predicted by GR. We
conclude that accurate determinations of this time delay effect that agree
with GR, using solar spacecraft probes, do not rule out the possibility of
detecting a deviation from GR in the geodetic spin precession due to an
NGT correction obtained from Eq.(\ref{NGTdeviation}).

\section{Conclusions}

We have derived the equation of motion for a spinning particle and the
geodetic precession of a gyroscope in the version of NGT published in
1995~\cite{Moffat4,Moffat5}. With the significant sensitivity of the
Gravity Probe-B experiment, it is hoped that the NGT correction to the
predicted GR gyroscope precession can be detected.

The time delay of a radio signal obtained from NGT agrees with the
prediction of GR. Therefore, an accurate determination of the time delay of
a radio signal from a spacecraft as it passes by close to the Sun, which
competes with the accuracy of the determination of the Gravity Probe-B
gyroscope's precession, does not rule out a detection of a correction to
the precession obtained from NGT. The predicted Lense-Thirring
frame-dragging effect in NGT is the same as in GR.

 \end{document}